\documentclass[lineno,authoryear]{FLO_v1}%

\usepackage{graphicx}
\usepackage{upgreek}
\usepackage{multicol,multirow}
\usepackage{amsmath,amssymb,amsfonts}
\usepackage{mathrsfs}
\usepackage{amsthm}
\usepackage[figuresright]{rotating}
\usepackage{appendix}
\usepackage[authoryear]{natbib}
\usepackage{ifpdf}
\usepackage[T1]{fontenc}
\usepackage{newtxtext}
\usepackage{newtxmath}
\usepackage{textcomp}
\usepackage{xcolor}
\usepackage{booktabs}
\usepackage{multirow}
\usepackage{threeparttable}
\usepackage{caption}

\usepackage[colorlinks,allcolors=blue]{hyperref}
\definecolor{jourcolor}{cmyk}{1,0.57,0.01,0.38}
\hypersetup{
    colorlinks,%
    citecolor=jourcolor,%
    filecolor=jourcolor,%
    linkcolor=jourcolor,%
    urlcolor=jourcolor
}

\theoremstyle{definition}

\articletype{RESEARCH ARTICLE}

\DOI{10.1017/flo.2026.1}

\Year{2021}

\Vol{1}

\Price{}

\art-id{FLO2000049}

\citearticle{McGlade, A. T., \& Buxton, O. R. H.}

\begin{document}

\title[Wind farm global blockage as an adverse pressure gradient problem: turbulence amplification and spectral modification in the induction region of a model wind farm]{Wind farm global blockage as an adverse pressure gradient problem: turbulence amplification and spectral modification in the induction region of a model wind farm}

\author{Adrian T. McGlade$^{1\ast}${\href{https://orcid.org/0009-0004-3748-3437}{\includegraphics{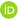}}} and Oliver R. H. Buxton$^{1}${\href{https://orcid.org/0000-0002-8997-2986}{\includegraphics{orcid_logo}}}}

\authormark{ATM and ORHB}

\address[1]{Department of Aeronautics, Imperial College London, UK}

\corres{*}{Corresponding author e-mail:
\emaillink{a.mcglade23@imperial.ac.uk}}

\keywords{Boundary Layers; Wind farms}

\date{\textbf{Received:} XX 2026; \textbf{Revised:} XX XX 2026; \textbf{Accepted:} XX XX 2026}

\abstract{
The induction region of a model wind farm is investigated experimentally as an adverse pressure gradient (APG) turbulent boundary layer problem, using hot-wire anemometry. An array of porous discs, with diameter $D=50$\,mm representing the farm, in multiple configurations across two turbulent boundary layers of different depths, with farm-present and farm-absent configurations compared throughout. The farm imposes a spatially developing adverse pressure gradient in the approach flow. Spanwise-averaged measurements reveal systematic farm-scale global blockage extending to at least $10D$ upstream in all cases. The farm-induced turbulence intensity increase, which reaches up to $7\%$ relative to the farm-absent reference for the shallowest boundary layer, is shown by cumulative variance decomposition to be carried predominantly by large-scale motions of the boundary layer, whilst the small-scale turbulence intensity is only weakly affected. This selective large-scale energisation is consistent with the outer-layer amplification driven by an adverse pressure gradient. The concentration of farm-induced turbulence energy at scales comparable to or larger than the rotor diameter has direct implications for turbine fatigue loading, preferentially exciting coherent rotor-scale load fluctuations at the frequencies most damaging under fatigue. The results establish a quantitative link between wind farm global blockage and APG boundary layer physics. 

}

\maketitle

\begin{boxtext}

\textbf{\mathversion{bold}Impact Statement}
This work establishes the upstream induction region of a wind farm as a physically interpretable adverse-pressure-gradient (APG) turbulent boundary layer problem, rather than a prescribed inflow. Farm-induced blockage is shown to systematically amplify large-scale motions in the outer layer, in direct analogy with canonical APG boundary layers, while the location of the energised scale \(\left( \lambda_x / \delta_{99} \approx 2-3\right)\) remains invariant, consistent with this scale reflecting the boundary layer's own receptive structure rather than one imposed by the farm. 
This work has direct relevance to wind-energy engineering. By identifying that farm blockage selectively energises turbulent scales comparable to or larger than the rotor diameter, the results imply that the induction region drives coherent, whole-rotor loading of the kind most damaging under fatigue, with turbine thrust and boundary layer depth being crucial parameters. These findings should be factored into future wind farm layout optimisation, to ensure that designs, particularly for the leading rows, are able to reliably meet their intended in-service lifetimes.

%This work establishes the upstream induction region of a wind farm as a physically interpretable adverse-pressure-gradient (APG) turbulent boundary layer problem, rather than a prescribed inflow. Using wind-tunnel experiments with two boundary layers of different depths and a model farm of porous discs, we show that farm-induced blockage systematically amplifies large-scale motions in the outer layer, in direct analogy with canonical APG boundary layers, while the location of the energised scale \(\left( \lambda_x / \delta_{99} \approx 2-3\right)\) remains invariant, consistent with this scale reflecting the boundary layer's own receptive structure rather than one imposed by the farm. 
%This work has direct relevance to wind-energy engineering. By identifying that farm blockage selectively energises turbulent scales comparable to or larger than the rotor diameter, the results imply that the induction region drives coherent, whole-rotor loading of the kind most damaging under fatigue, with turbine thrust and boundary layer depth being crucial parameters. 

\end{boxtext}
\newpage 
\section{Introduction}
The aerodynamic interaction between a wind farm and the atmospheric boundary layer (ABL) that drives it represents one of the more complex problems in environmental fluid mechanics. Most research on wind farm aerodynamics has focused on turbine wakes, the farm's upstream effect on the approaching boundary layer has received comparatively little attention. This upstream effect, termed global blockage \citep{Bleeg2018}, is a systematic deceleration of the incoming flow extending many rotor diameters ahead of the leading row, reducing available kinetic energy flux by 5--15\% for large arrays \citep{Segalini2020, Nishino2020}. The mechanism is established at the inviscid level: farm thrust imposes an adverse pressure gradient (APG) on the approach flow, elliptic in character and decaying on the scale of the farm rather than the rotor \citep{Burton2011, Segalini2020}, with deficits confirmed observationally \citep{Schneemann2021} and in RANS \citep{Bleeg2018} to scale with farm size rather than rotor size.

What remains largely unaddressed is how this farm-induced APG interacts with the turbulent structure of the turbulent boundary layer (TBL), despite APG-TBL interaction being one of the more thoroughly studied problems in classical fluid mechanics \citep{Clauser1954, Townsend1961, Bradshaw1967}. The induction region maps well onto this canonical problem: an APG preferentially decelerates the near-wall region relative to the outer layer, redistributing shear production outward and amplifying large-scale and very-large-scale motions (LSM/VLSM), the defining signature of APG boundary layers, the strength of the APG quantified by the Clauser parameter $\beta_C = (\delta^{*}/\tau_w)\,\mathrm{d}p/\mathrm{d}x$ where \(\tau_w\) is the wall shear stress and \(\delta^{\ast}\) is the displacement thickness \citep{Harun2013, Kitsios2016}. A key complication is that this turbulence state depends not only on local $\beta_C$ but on upstream pressure-gradient history \citep{Bobke2017, SanmiguelVila2017}: since the farm-induced gradient strengthens monotonically from near zero far upstream to a maximum at the first row, the boundary layer arriving at the farm carries an integrated history invisible to single-station measurements or to induction models that prescribe a fixed inflow profile. This forcing is also physically distinct from canonical laboratory APGs, which are typically constant or ramp-type, or are imposed via prescribed boundary conditions in numerical simulations; the farm's pressure field instead follows its own potential-flow solution, rising smoothly over a streamwise distance set by the farm geometry rather than any imposed tunnel or domain feature. The induction-region boundary layer is therefore never in Rotta--Clauser equilibrium, and instead closer to the weakly-to-moderately non-equilibrium regime studied by \citet{Harun2013} and \citet{Bobke2017}.

This outer-layer amplification is of practical concern as LSMs and VLSMs, with streamwise scales of $2$--$3$ and $10$--$20$ times the boundary layer depth respectively \citep{Hutchins2007, Kim1999}, carry a disproportionate share of outer-layer turbulent energy while also being comparable to or larger than modern rotor diameters, precisely the flow structures experienced by a rotor at ordinary hub heights. Such structures drive coherent, near-simultaneous loading across the whole rotor disc rather than loads relieved by spatial averaging, and \citet{Chamorro2012} showed that longer coherent inflow structures produce larger-amplitude, lower-frequency load cycles that are more damaging from a fatigue perspective \citep{Sutherland1999}. The farm-induced APG is therefore a physically distinct inflow problem from wake-added turbulence, with consequences for blade fatigue life \citep{Thomsen1999, Sutherland1999}, main shaft bearings \citep{Hahn2007, Nejad2014}, and tower base loading \citep{Murtagh2005, Frandsen2005}.

%Quantifying this response in practice is complicated by the reliance of canonical APG scaling on $u_\tau$ and $\delta^{*}$, both difficult to determine reliably in the present flow (see supplementary material for discussion); $\delta_{99}$ and $U_e = \bar{u}(z=\delta_{99})$ are therefore adopted throughout, following \citet{Harun2013}, \citet{Monty2011}, and \citet{Bobke2017}, and the Clauser parameter $\beta_C$ is consequently unavailable, so the boundary-layer response is characterised in terms of $C_T$ and $\delta_{99}/D$ rather than the conventional equilibrium-parameter framework.

\subsection*{Scope and aims of the present work}
The coupling between $C_T$ and blockage magnitude further suggests that the farm induction region and turbine operating state form a closed dynamical system even before this turbulence mechanism is considered: \citet{Bleeg2018} and \citet{Bleeg2022} showed blockage depends directly on collective thrust loading rather than turbine geometry, \citet{Delvaux2025} showed farm power/thrust curves and the optimal collective set point shift under blockage in a way that requires farm thrust and the upstream pressure field to be mutually determining, and \citet{Bossanyi2024} showed axial-induction control can reduce blockage and wake losses simultaneously. \citet{Chamorro2012}, \citet{Chatterjee2018} and \citet{Gambuzza2021} demonstrated the importance of large-scale coherent inflow structures for turbine loading and wind-farm power, while \citet{Revaz2025} showed that increased turbulence can modify turbine induction, collectively indicating that amplified large-scale turbulence may influence the effective aerodynamic thrust of a wind farm. What these studies do not address is the effect of the farm-generated APG on the turbulent structure of its own inflow: amplified large-scale turbulence modifies the effective thrust of the farm, in turn modifying the upstream pressure gradient, closing a feedback loop implicit in the $C_T$--blockage literature but not previously characterised in its turbulent boundary layer component.

The present work experimentally isolates this boundary layer component, in the absence of the thrust feedback that would close the full loop. The farm is represented by an array of porous discs, shown to reproduce the time-averaged turbine and farm-scale pressure field and upstream momentum deficit \citep{Bossuyt2018, Vinnes2023, Gouder2024, Ahmed2024} while suppressing the dynamic thrust response, blade-passage forcing, and tip-vortex turbulence of real rotors, allowing the APG-driven outer-layer amplification to be characterised without the spectral influence of rotor dynamics. Two boundary layers of different depths are examined for matched farm geometries, isolating the boundary depth. Farm-present and farm-absent configurations are compared under matched conditions throughout to separate the farm-induced response from the natural streamwise development of the boundary layer.

\section{Experimental Setup}
% ================================================================== %

The experimental campaign was undertaken in Imperial College London's John Harvey wind tunnel to investigate the induction region of a model wind farm and its interaction with the approaching turbulent boundary layer. The John Harvey wind tunnel is a closed-loop facility with a working section $4000\,\text{mm}$ long, $1650\,\text{mm}$ wide and $1100\,\text{mm}$ tall. The experiment was conducted at $U_\infty \approx 19.5\,\mathrm{m}/\mathrm{s}$. %The reference velocity $U_\text{ref}$ was taken at $(x/D,\,y/D,\, z/D) = (-8,\,0,\,10)$ relative to the centre of the leading row of discs for all cases.
%($80D \times 33D \times 22D$ for a disc diameter $D = 50\,\text{mm}$)

Single-component hot-wire anemometry was used to characterise the flow over a spatial domain capturing the induction region. The farm is represented by an array of porous discs (disc diameter $D = 50\,\text{mm}$), with multiple farm configurations being investigated. An illustration of the experimental set-up is shown in Figure 1, defining the coordinate system used throughout this work: $x$~streamwise, $y$~spanwise, and $z$~wall-normal. 

\begin{figure}[ht]
    \centering{\includegraphics[width=\textwidth]{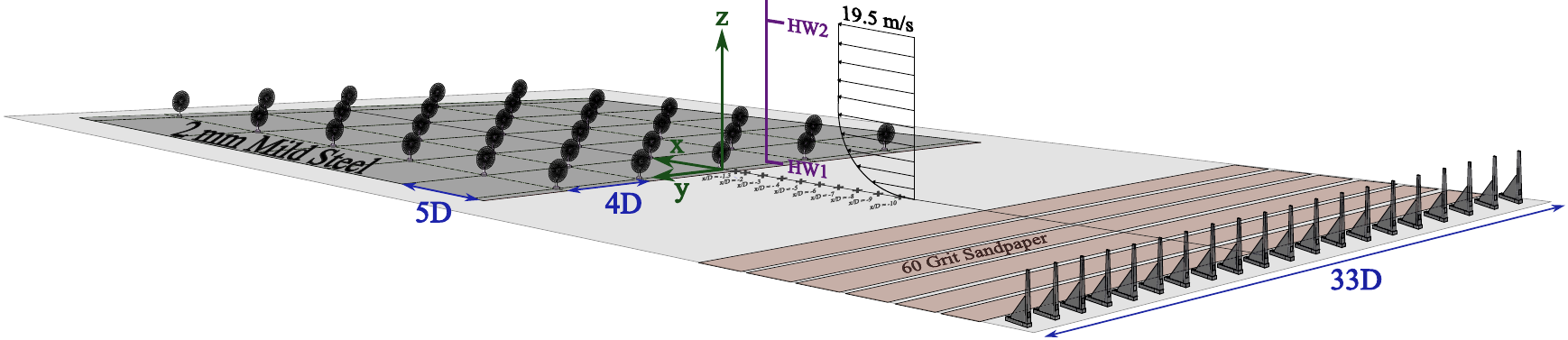}}
    \caption{Illustration of the experimental set-up and the coordinate system used throughout}
    \label{fig_layout}
\end{figure}

\begin{figure}[ht]
    \centering{\includegraphics[width=0.8\textwidth]{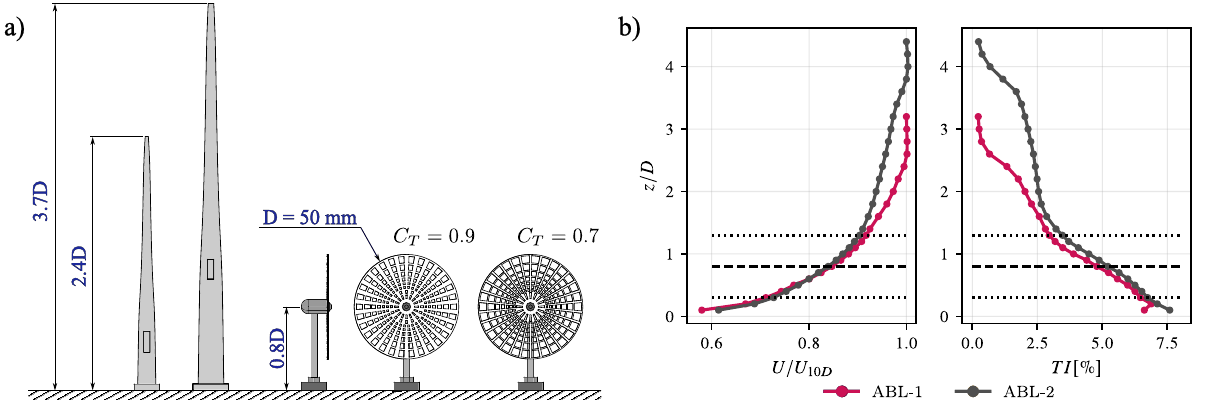}}
    \caption{a) Scale illustrations of the implemented spires to produce ABL~1 and ABL~2 with the \(C_T = 0.9\) and \(C_T = 0.7\) disc assemblies; and b) incoming boundary layer profiles for the ABL cases at $x/D = 0$, with the farm absent: normalised mean velocity $U/U_\text{10D}$, turbulence intensity, TI. The hub-height is marked by the bold dashed line, with the dotted black lines indicating the extent of the disc.}
    \label{fig_layout_V2}
\end{figure}

% ==================================================================== %
\subsection{Velocity measurement}\label{sect:vel_measure}
% ==================================================================== %

The flow was measured using hot-wire anemometry: two probes sampled simultaneously, both attached to an \(xy\) traverse. Probe positioning error was $< 0.02D$. The primary probe (HW1) was also traversed in the wall-normal direction, while the secondary probe (HW2) was fixed at \(z/D = 10\) to record the freestream velocity. Both hot wires were Dantec Dynamics 55P11 single-wire probes ($5\,\mu\text{m}$ diameter, $1.25\,\text{mm}$ sensing length), mounted in Dantec 55H26 right-angle and 55H21 straight supports and operated in constant-temperature mode using Dantec 91C10 CTA modules connected to a Dantec StreamLine 90N10 frame. Signals were sampled at $10\,\text{kHz}$. Single-wire probes of this type are well established for turbulence and boundary layer measurements \citep{Hutchins2009, Ligrani1987}, providing high temporal resolution and good signal-to-noise ratio.

Hot-wire voltages were converted to velocity following \citet{Bourhis2024}, using fourth-order polynomial fits with the temperature correction of \citet{Hultmark2010}:
\begin{equation}%\label{eq:HW}
    \frac{U}{\nu} = f_4\!\left(\frac{E^2}{k\,\Delta T}\right),
    \qquad f_4 \text{ a fourth-order polynomial,}
\end{equation}
where $\nu$ is the kinematic viscosity and $k$ the thermal conductivity, estimated by the methods of \citet{Smits2005} and \citet{Kannuluik1951}, respectively. The probes were calibrated against a Pitot-static tube connected to a Furness Control FCO560 micro-manometer approximately every $12$ hours or before and after runs (whichever came first); calibration coefficients were linearly interpolated in time between successive calibrations. The freestream temperature varied by at most $0.8^\circ\text{C}$ over the longest ($\approx 12\,\text{h}$) sessions, and post-correction velocity variability across calibrations was $<2\%$.

Two important limitations of this approach are noted. First, the probes and probe supports are physically intrusive; although their influence on the freestream is small, their potential effect on the near-wall region and on the induction signal cannot be entirely eliminated. Second, the probe geometry precludes reliable measurements very close to the wall: the finite probe length and support geometry prevent positioning below $z/D = 0.1$, so near-wall quantities such as the friction velocity $u_\tau$ cannot be determined directly.

The principal quantities of interest are the normalised mean velocity and turbulence intensity. Following the Reynolds decomposition:
\begin{equation}%\label{eq:Reyn_decomp}
    u(x,y,z,t) = U(x,y,z) + u'(x,y,z,t),
\end{equation}
$U$ is the temporal mean and $u'$ is the fluctuation. From common wind energy practice, these are expressed as the normalised mean wind speed $U/U_\text{ref}$ and turbulence intensity $\mathrm{TI} = \sqrt{\overline{u'^2}}/U_\text{ref}$.

The definition and measurement location of $U_\text{ref}$ in the wind farm literature has not reached consensus. \citet{Segalini2020} measured $U_\text{ref}$ approximately $8D$ upstream of the front row; \citet{Vinnes2023} define their reference near the tunnel roof, without specifying an exact streamwise position, but note no measurable sensitivity to $x/D$. Some authors instead use (more local) hub-height velocities, reflecting their focus on farm power (\citet{Bossuyt2017}; \citet{Bossuyt2018}) and wake development (e.g. \citet{Messmer2025}) rather than the upstream induction region itself. For the present work, resolving the spatial structure of the induction region requires a reference velocity taken outside that region, ruling out a local hub-height definition; $U_\text{ref}$ is therefore taken at a fixed upstream station, $(x/D,\,y/D,\,z/D) = (-8,\,0,\,10)$, similar to \citet{Segalini2020}. This is also the location at which is also the location at which the hot-wires were calibrated.

% ========================================================================= %
\subsection{Measurement strategy}
% ========================================================================== %

%The principal quantities of interest are the normalised mean velocity $U/U_\text{ref}$ and turbulence intensity TI. 
Measurements were split into two parts. Boundary layer sweeps in the \(xz\)-plane ($y=0$) for each streamwise position \(x/D\in [-10,\,-1.3]\), to evaluate the evolution of the boundary layer in the induction region. Samples were 120 seconds in duration. Secondly, sweeps in the $xy$-plane at the hub-height $z/D = 0.8$, covering $(x/D,\,y/D) \in [-10,\,-1.3] \times [-Y_F-2,\, Y_F+2]$ with typical resolutions of $\Delta x/D \le 1$, $\Delta y/D = 0.5$. Here, \(\mathrm{F}_Y\) is the farm half-width. Samples were 60 seconds in duration. Following \citet{Nishino2020} and \citet{Messmer2025} for the hub-height sweeps, spanwise averaging is applied to extract the effective farm-scale inflow, representing the appropriate quantities to analyse global blockage and the farm-scale momentum balance:
\begin{equation}%\label{eq:int}
    \langle \xi \rangle_y =
    \frac{1}{y_1 - y_0}
    \int_{y_0}^{y_1} \xi(x,y,z)\,\mathrm{d}y,
\end{equation}
where $\xi$ denotes $U$ or TI, and $(y_0/D,\,y_1/D) = (-\mathrm{F}_Y-2,\,\mathrm{F}_Y+2)$.

The measurement strategy therefore serves two complementary purposes: the $xz$-plane sweeps resolve the wall-normal boundary-layer structure, while the hub-height sweeps provide the spanwise-averaged, farm-scale inflow. These measurements do not, however, permit direct application of conventional APG boundary-layer scaling. The near-wall measurement limitations identified in Section~\ref{sect:vel_measure} (expanded upon in the supplementary material) preclude a reliable estimate of \(u_\tau\), while \(\delta^{*}\) requires a well-resolved near-wall profile and an unambiguous reference velocity, neither of which is available in the farm-modified induction region. We therefore, use $\delta_{99}(x)$, defined as the wall-normal location where $U_{99}(x) =  U(x,\delta_{99}) = 0.99\, U_\text{10D}(x)$, following \citet{Harun2013}, \citet{Monty2011}, and \citet{Bobke2017}), and characterise the boundary-layer response in terms of \(C_T\) and \(\delta_{99}/D\) rather than the conventional Clauser parameter \(\beta_C\).  
%This reflects the competing requirements of resolving the farm-scale response through spanwise averaging and resolving the boundary-layer structure through near-wall measurements.

% ------------------------------------------------------------------ %
\subsection{Wind farm model}
% ------------------------------------------------------------------ %

Porous discs serve as the experimental analogue of the actuator disc, with the disc drag coefficient playing the role of the turbine thrust coefficient $C_T$ in governing the upstream pressure rise and wake behaviour. The use of porous discs to model wind turbines alone and in farm configurations is well established \citep{Aubrun2013, Camp2016, Howland2016, Bossuyt2017, Stevens2017}. \citet{Aubrun2013} showed that the mean velocity and turbulence intensity of a porous disc wake agree with those of a rotating model turbine beyond $3D$ downstream; \citet{Stevens2017} confirmed this finding through comparison of LES with wind tunnel data and demonstrated that the porous disc faithfully reproduces the main features of the mean wake. There has been a progressive shift in the literature from uniform porous discs \citep{Aubrun2013} toward non-uniform porosity distributions \citep{Camp2016, Howland2016, Helvig2021}, motivated by various improvements in emulation of a three-bladed rotor \citep{Camp2016} and better reproduction of the physical length scales of the wake \citep{Helvig2021}. The use of porous discs is not without limitations: \citet{Aubrun2019} and \citet{Vinnes2022} showed that wake behaviour can be sensitive to initial conditions and disc design, particularly in the near wake and for higher for higher-order flow statistics. However, \citet{Vinnes2023} demonstrated that within wind farm arrays, porous discs are more faithful than for a single turbine, as the flow converges to similar statistics after the second row regardless of model type. In the context of global blockage specifically, porous discs are established in experimental studies \citep{Bossuyt2018, Vinnes2023}, and \citet{Gouder2024} showed that a porous disc can faithfully reproduce the induction zone of a model turbine. For the induction region the porous disc is therefore an appropriate model. 

Two disc designs of differing porosity were used, both modified versions of \citet{Bourhis2024}'s design, itself based on \citet{Camp2016}. The discs were laser-cut from $0.8\,\text{mm}$ stainless steel sheet. The \citet{Camp2016} design was adapted to ensure all features exceed $0.5\,\text{mm}$ (the minimum reliable dimension for the laser cutter used) whilst preserving the total porosity, giving $C_T = 0.7$ and $C_T = 0.9$. The $C_T$ of each disc was determined from a separate hydrodynamic flume experiment using load-cell and PIV data at $\mathrm{Re}_D = U_{hub} D/\nu \approx 5 \times 10^4$; \citet{Camp2016} and \citet{Messmer2025} report the discs to be $\mathrm{Re}_D$-independent for $\mathrm{Re}_D > 10^4$. 
Each disc is mounted to a 3D-printed PLA nacelle on a $3\text{~mm}$ diameter stainless steel mast. This configuration maintains a $D/d_{\text{mast}}$ ratio and a hub height of $H_{\text{hub}}/D = 0.8$, both matching the NREL $15\text{~MW}$ reference turbine geometry \citep{NREL}. The nacelle design allows for simple changing of the discs, and thus $C_T$. The mast is threaded into a $12\,\text{mm}$ diameter neodymium magnet which attaches to a 2mm mild steel base plate, allowing rapid repositioning and configuration changes. Though practically required, the inclusion of the nacelle and mast is also supported by the findings of \citet{Stevens2017}, whose simulations showed that their inclusion substantially improves wake fidelity relative to an isolated actuator disc. The discs and towers are illustrated in figure~\ref{fig_layout_V2}a).

Multiple farm configurations were considered, detailed in \hyperref[tab:BL_farm]{Table 1}, with principal streamwise and spanwise spacings of $S_x = 5D$ and $S_y = 4D$.  The choice of $S_x = 5D$ and $S_y = 4D$ represents an intermediate density between the dense arrays ($S_x, S_y \approx 3D$) of \citet{Segalini2020} and typical full-scale spacings of $S_x\approx 7-10D$ \citep{Meyers2012}, and is known to produce measurable global blockage \citep{Segalini2020, Vinnes2023}. However, these spacings are independently changed to evaluate their impact. The default farm configuration is 5 rows (streamwise count) and 5 columns (spanwise), with rows and columns being added and removed to investigate the impact on the farm induction region. The inclusion and removal of columns are of practical note in the hub-height sweeps as the width of the farm changes. The spanwise extent of the sweeps is thus the farm width \(\mathrm{F}_y\pm2D\), to not bias the results by including or excluding the lateral by-pass of flow around the farm. 
The farm is ordinarily aligned, with one disc sitting directly downstream of the preceding disc. Staggered cases are also considered, in which discs in each row are offset laterally to sit between the wakes of the preceding row. Lastly, the freestream velocity is reduced to \(U_{\infty} \approx 10 \mathrm{m}/\mathrm{s}\) to investigate any Reynolds number effects. 

\begin{table}[ht]  % Table 1
    \centering
    \begin{threeparttable}
        \caption{Implemented farm configurations.}
        \label{tab:BL_farm}
        
        \begin{tabular}{ccccccc}
            \toprule
            ABL & $C_T$ & Rows & Columns & Total Discs & Config. Name & Notes \\
            \midrule
            \multirow{11}{*}{ABL 1} 
                & 0.7 & 4 & 5 & 20 & 4x5 & \\
                &     & 4 & 6 & 24 & 4x6 & \\
                &     & 5 & 5 & 25 & 5x5 & \\
                &     & 5 & 5 & 25 & 5x5-Stag & Staggered \\
                &     & 6 & 4 & 24 & 6x4 & \\
                &     & 6 & 5 & 30 & 6x5 & \\
                \cmidrule(lr){2-7}
                & 0.9 & 4 & 5 & 20 & 4x5 & \\
                &     & 4 & 6 & 24 & 4x6 & \\
                &     & 5 & 5 & 25 & 5x5 & \\
                &     & 6 & 4 & 24 & 6x4 & \\
                &     & 6 & 5 & 30 & 6x5 & \\
            \midrule
            \multirow{9}{*}{ABL 2} 
                & 0.7 & 5 & 5 & 25 & 5x5 & \\
                \cmidrule(lr){2-7}
                & 0.9 & 3 & 5 & 15 & 3x5 & \\
                &     & 4 & 5 & 20 & 4x5 & \\
                &     & 5 & 5 & 25 & 5x5 & \\
                &     & 5 & 5 & 25 & 5x5-Sp5  & $S_X = 5$ \\
                &     & 5 & 5 & 25 & 5x5-Str7 & $S_Y = 7$ \\
                &     & 6 & 5 & 30 & 6x5 & \\
                &     & 7 & 5 & 35 & 7x5 & \\
                &     & 6 & 5 & 30 & 6x5-10 & \(U_{\infty} = 10\mathrm{m}/\mathrm{s}\) \\
                &     & 6 & 5 & 30 & 6x5-10-Stag & \(U_{\infty} = 10\mathrm{m}/\mathrm{s}\),  Staggered \\
            \bottomrule
        \end{tabular}

        \begin{tablenotes}
            \small
            \item Default grid spacing is $S_X = 5, S_Y = 4$ and aligned, unless specified in the Notes.
        \end{tablenotes}
    \end{threeparttable}
\end{table}

% ================================================================= %
\subsection{Boundary layer generation}
% ================================================================== %

Spires and surface roughness were used to generate ABL-like turbulent boundary layers at controlled depths. Spires were initially designed following \citet{Irwin1981}; a second generation of spires incorporated the modifications proposed by \citet{Hobson2015}, who refined the triangular spire profile through iterative parametric CFD and wind tunnel studies to improve boundary layer fidelity at small scales. These modifications alter the vertical variation of spire width, changing how the spire interacts with the flow at different wall-normal positions and thereby reshaping the generated boundary layer profile. 

Two spire geometries were used, illustrated in figure~\ref{fig_layout_V2}a), a smaller spire (of spire height, $H = 120\, \text{mm} = 2.4D$) generates ABL~1. A second, taller spire ($H = 185\text{mm} = 3.7D$) generates ABL~2. Both incorporate the modifications of \citet{Hobson2015} to the Irwin profile, and the fetch exceeds the minimum $6H$ recommended by \citet{Irwin1981} for both . Due to the small scale of the experiment, discrete roughness elements were impractical; surface roughness was instead applied as 60-grit sandpaper over a $1250\,\text{mm}$ fetch, followed by a $250\,\text{mm}$ smooth recovery region and a $500\,\text{mm}$ smooth wall induction region before the leading disc row. The spire bases are $2000\,\text{mm}$ upstream of the leading disc row, and a string trip $100\,\text{mm}$ upstream of the spires fixes the boundary layer origin. The spires were laser-cut from $6\,\text{mm}$ cast acrylic; their tips were truncated to 90\% of the nominal height to avoid localised laser heating artefacts near sharp features.

The two boundary layers were designed to match hub-height mean velocity, turbulence intensity, and integral length scale $L_u$ (via Taylor's hypothesis), whilst differing in boundary layer depth. This isolates the effect of $\delta_{99}$ on the farm induction interaction as cleanly as practical, though a pure isolation was not possible: achieving different boundary layer depths with matched hub-height statistics necessarily produces differences in the wall-normal profile shape, characterised here by the Hellman exponent $\alpha$ from a power-law fit with reference height $z_\text{ref} = z_{hub}$. Due to its lower depth, ABL~1 produces greater shear. The resulting boundary layer profiles for both ABL cases are shown in Figure~\ref{fig_layout_V2}b), with key parameters summarised in Table~\ref{tab:BL}. 

\begin{table}[ht] % Table 2
    \centering
    \begin{threeparttable}
        \captionsetup{width=\textwidth}
        \caption{Properties of the generated ABL-like inflows at $x/D = 0$.}
        \label{tab:BL}
        
        \begin{tabular}{lccc}
        \toprule
         & ABL~1 & ABL~2 \\
        \midrule
        Mean velocity, $U/U_\text{ref}$ at \(z_{hub}\)
            & 0.847 & 0.839 \\
        Turbulence intensity, TI [\%] at \(z_{hub}\)
            & 4.84  & 5.21   \\
        Integral length scale, $L_u/D$ at \(z_{hub}\)
            & 1.1 & 1.2  \\
        Boundary-layer thickness, $\delta_{99}/D$
            & 2.4 & 3.6   \\
        Hellman exponent, \(\alpha\) 
            & 0.156 & 0.128 \\
        Friction velocity\(^*\), $u_\tau [\text{m.s}^{-1}]$
            & 0.6  & 0.6 \\
        \(\mathrm{Re}_{\tau}\)\(^*\),
            & 4800 & 7700 \\
        Pressure gradient\(^*\),
            & $\frac{\partial P}{\partial x} > 0$
            & $\frac{\partial P}{\partial x} > 0$ \\
        \bottomrule
    \end{tabular}

        \begin{tablenotes}
            \small
            \item \(^*\) Indicative estimates only.
        \end{tablenotes}
    \end{threeparttable}
\end{table}

\section{Results and Discussion}
The full spatial structure of the upstream flow modification in $yz$-planes are presented in the supplementary material, showing a local, disc-scale velocity deficit that decays by $x/D \approx -3$ superposed on a broader, spanwise-uniform farm-scale deceleration and turbulence response; this motivates the spanwise-averaged and boundary-layer focus of the remainder of this work.
% ------------------------------------------------------------------ %
\subsection{Boundary layer structure and pressure gradient}
%\label{sec:BL_profiles}
% ------------------------------------------------------------------ %

\begin{figure}[ht] % Figure 3
    \centering
    \includegraphics[width=8cm]{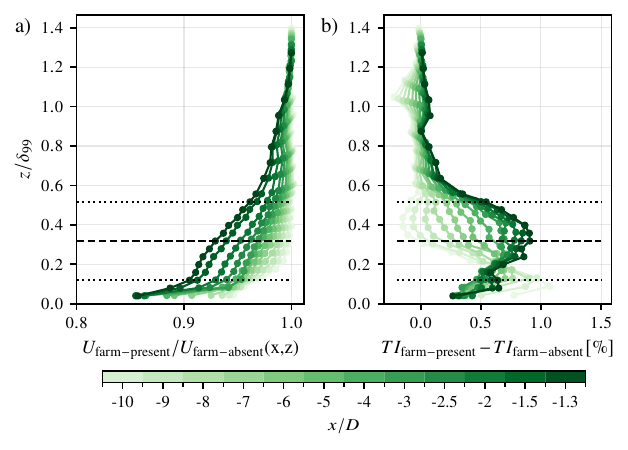}
    \caption{a) Velocity ratio and b) TI difference for ABL~1.} 
    \label{fig_bl_defect} 
\end{figure}

Figure~\ref{fig_bl_defect} presents wall-normal profiles of $U_\mathrm{farm-present}/U_\mathrm{farm-absent}$ and $\mathrm{TI}_{\mathrm{farm-present}} - \mathrm{TI}_{\mathrm{farm-absent}}$ at the farm centreline ($y/D = 0$) for successive streamwise stations from $x/D = -10$ to $x/D = -1.3$, for both the farm-present and farm-absent configuration of ABL~1. The farm is comprised of five rows and five aligned columns with the \(C_T = 0.7\) discs, as an illustrative example. The streamwise velocity $U/U_\text{ref}$ is modified throughout the measurement domain in the farm-present case. Two features are immediately apparent. The first is the strong local modification near hub height in the vicinity of the leading disc row, attributable to individual disc induction. The second, and more significant for the present purposes, is a systematic reduction in the near-wall velocity that extends considerably further upstream than the disc-scale effect (and far beyond what a hub-height only investigation would reveal) and is consistent with the preferential deceleration of the low-momentum near-wall fluid under an adverse pressure gradient - precisely the behaviour documented in classical APG boundary layer studies \citep{Clauser1954, Townsend1961}. This upstream near-wall modification also underlines the difficulty of using inner scaling in the induction region, since the wall shear stress is itself being modified by the farm pressure field at stations far upstream of the farm. Figure~\ref{fig_bl_defect}b) shows the difference in TI due to the presence of the farm. Notable is the increase in TI in the vertical extent of the disc. \(z_{hub}/\delta_{99} \approx 0.3\), which is close to the usual \(z\)-position of the outer peak of an APG boundary layer as shown by \citep{Harun2013}. This has two important consequences for wind farms / wind turbines, with typical hub heights of modern turbines \(\mathcal{O}(100\mathrm{m})\) \citep{NREL} and ordinary offshore boundary layer depths in the range of \(200-800\mathrm{m}\) \citep{Krogsaeter2015}, the rotor will sample the LSMs of the boundary layer. Secondly, the presence of the the farm through a farm induced pressure gradient may enhance these LSMs. 

For ABL~1 and ABL~2, direct pressure measurements are not available and the character of the approach flow pressure gradient must be inferred from the streamwise evolution of the mean velocity. In the farm-absent configuration, both cases exhibit a mild adverse pressure gradient throughout the measurement domain, evidenced by a weak but consistent decrease in $U$ with increasing $x/D$ at all wall-normal heights. The introduction of the farm enhances this gradient: $U$ decreases monotonically as the leading disc row is approached, with the deceleration intensifying toward the farm face. 

\subsection{Farm-scale induction: spanwise-averaged velocity and turbulence intensity}
\label{sec:spanwise}
% ------------------------------------------------------------------ %

\begin{figure}[ht] % Figure 4
    \centering
    \includegraphics[width=12cm]{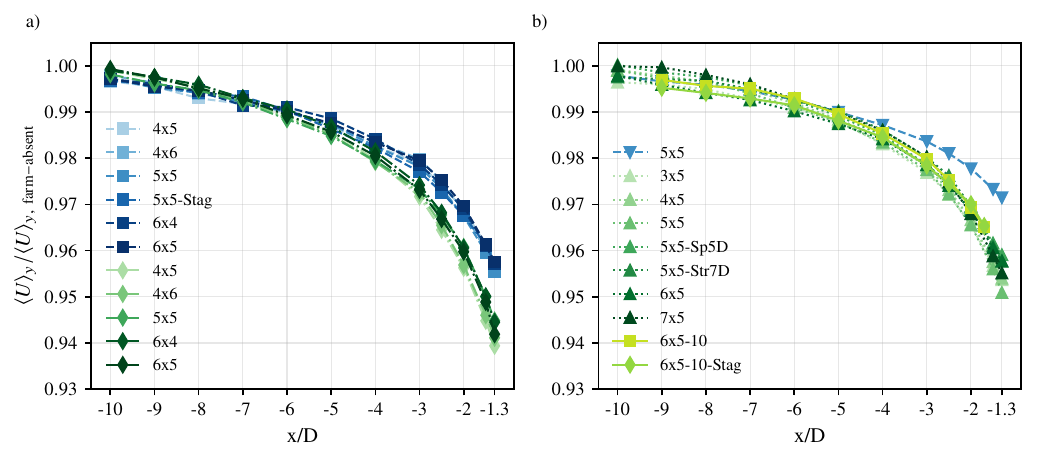}
    \caption{Hub-height spanwise averaged mean velocity \(U\) for a) ABL~1 and b) ABL~2. All normalised by the farm-absent case. \(C_T = 0.7\) in shades of blue, \(C_T = 0.9\) cases in in shades of green.} 
    \label{fig_compare_all_vel}
\end{figure}

Figure~\ref{fig_compare_all_vel} presents the spanwise-averaged normalised hub-height mean velocity $\langle U\rangle_y / \langle U \rangle_{y,\ \mathrm{farm-absent}}$ as a function of streamwise position, for both ABLs and all farm configurations.  A systematic velocity deficit relative to the farm-absent case is observed in all configurations across the full measurement domain, confirming the presence of farm-scale global blockage. The deficit grows monotonically as the leading disc row is approached and the farm-induced pressure rise intensifies. Measurements extend to \(x/D = -10\), ABL~1 and ABL~2 are only beginning to approach unity at this stage, indicating that the farm-scale induction region extends at least \(10D\) upstream in both cases. These observations are consistent with the results of \citet{Schneemann2021}, who identified upstream velocity deficits at distances of $5-20\text{km}$ ahead of large North Sea farms, and with the RANS predictions of \citet{Bleeg2018}

ABL~2 shows a weaker deficit than ABL~1 at equivalent upstream stations and approaches unity more rapidly, suggesting that the deeper boundary layer recovers toward the undisturbed state over a shorter normalised distance. The inability to identify a clean upstream onset within the measurement domain in any of the cases also cautions against interpreting the velocity ratio at a single upstream station as a measure of total blockage magnitude, the apparent deficit depends on the choice of reference station, and a reference taken within the induction region will systematically underestimate the true farm-induced velocity deficit. The ranking ABL~1 $>$ ABL~2 is consistent with the expectation that a shallower boundary layer constrains the available vertical bypass and therefore intensifies the farm-scale blockage, a result in qualitative agreement with the two-scale momentum theory of \citet{Nishino2020}, the simulations of \citet{Ivanell2026} and the stratification-dependent blockage documented by \citet{Bleeg2022}. 

Numerous farm configurations were considered, and seemingly for the configuration implemented, they show little impact on the farm blockage. For both ABLs, (comparing 4 and 6 rows in ABL~1, or 3 and 7 rows in ABL~2) increasing rows or changing streamwise and spanwise spacing or aligned and staggered shows little impact on the farm-scale blockage. This somewhat paradoxical result is consistent with the work of \citet{Segalini2020}, who showed that farm blockage is quickly saturated, since the first three rows contribute the most, and subsequent rows having a minor impact. The relation with the disc thrust coefficient is markedly different, the configurations with \(C_T = 0.9\) produce more blockage than those with \(C_T = 0.7\), consistent across both ABLs. An interesting conclusion in an operational context. If the first few rows are so impactful in terms of global blockage, and blockage being closely tied to turbine loading, could derating the leading rows meaningfully mitigate blockage? These results do not appear to be highly dependent on \(\mathrm{Re}_D\), the freestream velocity was reduced to \(10\mathrm{m}/\mathrm{s}\) and the results are presented, for ABL~2 with a \(6 \times 5\) farm with \(C_T =0.9\) discs, showing very good adherence to the results with double the Reynolds number. 

\begin{figure}[ht] % Figure 5
    \centering
    \includegraphics[width=12cm]{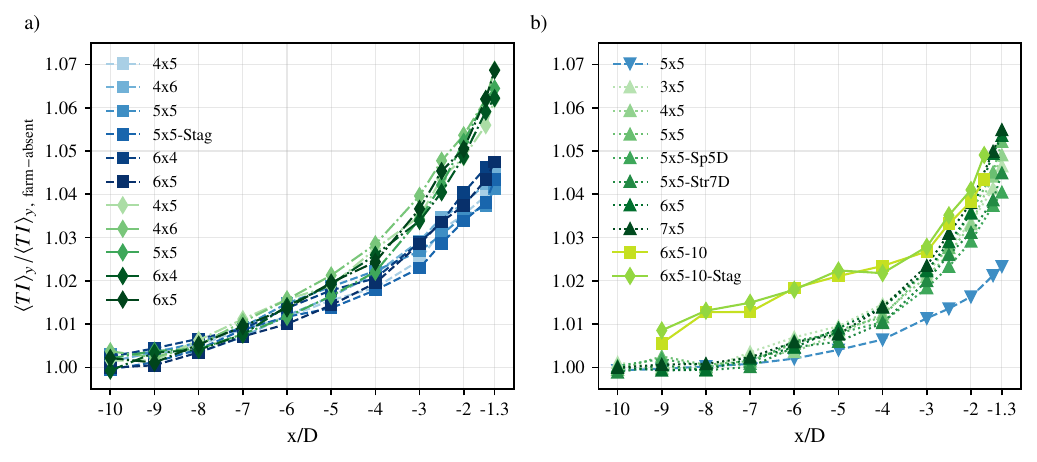}
    \caption{Hub-height spanwise averaged TI for a) ABL~1 and b) ABL~2. All normalised by the farm-absent case. \(C_T = 0.7\) in shades of blue, \(C_T = 0.9\) cases in in shades of green.} 
    \label{fig_compare_all_TI}
\end{figure}

The spanwise-averaged turbulence intensity ratio $\langle \mathrm{TI} \rangle_y / \langle \mathrm{TI} \rangle_{y,\ \mathrm{farm-absent}}$ is shown in Figure~\ref{fig_compare_all_TI}. Like the velocity ratio, the TI ratio exhibits a monotonically ordered response for both ABL cases: ABL~1~$>$~ABL~2. This ordering is consistent with the expectation that a shallower boundary layer, in which the farm pressure field acts over a larger fraction of the boundary layer depth relative to the energetic outer-layer scales, drives a more intense APG response in the outer layer. The boundary layers in the farm-absent configuration already exhibits a mild adverse pressure gradient ($\partial U/\partial x < 0$ at hub height), so the boundary layer arrives at the farm face with a pre-existing APG history. The introduction of the farm enhances this gradient, but the boundary layer is already adjusting toward an APG state. The pressure gradient memory of the boundary layer \citep{Bobke2017, SanmiguelVila2017} means that boundary layers subjected to the same local $\beta_C$ but different upstream histories present differently: \citet{Bobke2017} and \citet{SanmiguelVila2017} showed that flows with an increasing $\beta_C$ history retain some character of their (weaker) upstream condition, exhibiting behaviour closer to a lower-$\beta_C$ state than the local value alone would suggest, whereas flows with a decreasing $\beta_C$ history retain the imprint of a stronger upstream condition and show correspondingly elevated Reynolds stresses. The farm induction region is one of increasing-$\beta_C$ history, strengthening monotonically from a small value far upstream far upstream to its maximum at the leading rotor row; on this basis, the boundary layer arriving at the farm face may be expected to under-respond relative to what its local $\beta_C$ alone would predict, an effect that cannot be quantified from local measurements alone and is a further source of ambiguity in relating the present flow to equilibrium APG scaling.

Similarly to the mean velocity blockage, the TI response seems to be little influenced by the character of the farm. However, as for velocity, the effect of \(C_T\) is clear. Moving from \(CT=0.7\) to \(0.9\) results in a marked increase in TI amplification. As shown by \citet{Ahmed2024}, the strength of a disc's induction region increases with increasing \(C_T\), we can conclude that the APG induced by the farm is enhanced by increasing the \(C_T\) of the discs. It would be interesting to investigate if the TI amplification reaches the same degree of saturation seen in the velocity blockage (as seen in \citet{Segalini2020}), and whether derating the leading rows could meaningfully mitigate the TI amplification. % of a large farm. 

A noteworthy feature of the TI ratio in Figure~\ref{fig_compare_all_TI} is that measurements are limited to \(x/D \leq -1.3\) due to the hot-wire probe support, meaning we cannot comment on the turbulence immediately in front of the farm face. The enhancement may continue or possibly transition toward a rapid distortion theory (RDT) regime very close to the farm face, where the strain rate imposed by the farm pressure field becomes sufficiently large relative to the turbulent timescales that the turbulence cannot respond through the usual production-dissipation mechanisms and instead undergoes rapid distortion \citep{Hunt1990, Savill1987}. In the RDT limit, an irrotational straining field suppresses the streamwise Reynolds stress component $\overline{u'^2}$ whilst amplifying the cross-stream components, producing a net reduction in TI as conventionally defined. Direct experimental evidence for RDT behaviour in the near-disc induction zone has been reported by \citet{Gouder2024}, who identified suppression of the streamwise turbulence component immediately upstream, in the same region identified here, of a porous disc model turbine consistent with the RDT predictions. Initial investigations with a laser Doppler velocimetry system, show a decrease in TI in the region of \(x/D \approx -0.8\) pointing towards an RDT-like effect. 

The consistent ranking of mean velocity blockage and turbulence intensity amplification across the ABL cases and farms is a central finding of the present work. For the mean flow, this ranking is consistent with the degree of confinement of the farm within the inflow boundary layer: a shallower boundary layer places the farm within a proportionally larger fraction of the layer's own depth, restricting the bypass geometry available for flow to divert over and around the array and thereby amplifying the pressure-driven mean-flow deceleration, consistent with the established dependence of blockage on boundary layer depth reported in \citep{Ivanell2026, Nishino2020}. The turbulence intensity response follows the same ranking, and confinement offers a plausible extension of this mechanism: a farm occupying a larger fraction of $\delta_{99}$ represents a proportionally larger perturbation to the boundary layer's own outer-scale motions, which could drive stronger energisation of the outer-layer large-scale structures. However, the two boundary layers examined here likely also differ in upstream pressure-gradient history. Under the turbulent memory effect \citep{Bobke2017, SanmiguelVila2017}, this history alone could account for part of the observed difference in turbulence amplification, independent of any confinement mechanism. The present dataset cannot separate these two contributions, and the consistent ABL~1 $>$ ABL~2 ranking observed for the turbulence response should therefore be interpreted as consistent with $\delta_{99}/D$ acting as an important parameter, rather than as evidence that confinement alone drives the turbulence ranking.

% ------------------------------------------------------------------ %
\subsection{Spectral energy maps}
\label{sec:spectra_maps}
\begin{figure}[ht] % Figure 6
    \centering
    \includegraphics[width=12cm]{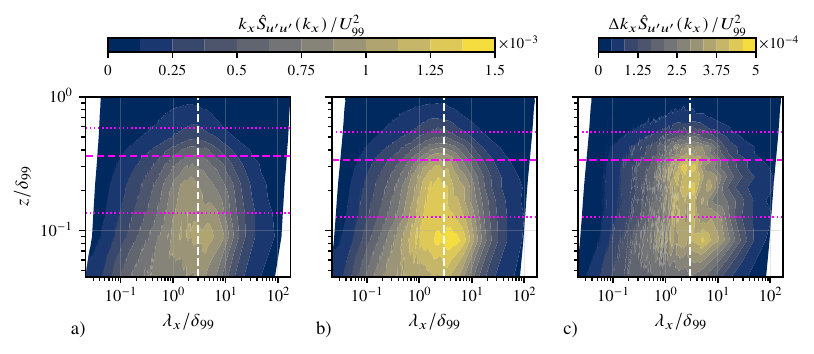}
    \caption{Normalised spectral energy map for ABL 1 at \(x/D = -1.3\) for a) farm-absent and b) 5x5 farm with \(C_T=0.9\) discs and c) difference. Dashed magenta lines indicate the hub-height and vertical extent of the disc}
    \label{fig_em_100}
\end{figure}

\begin{figure}[ht] % Figure 7
    \centering
    \includegraphics[width=12cm]{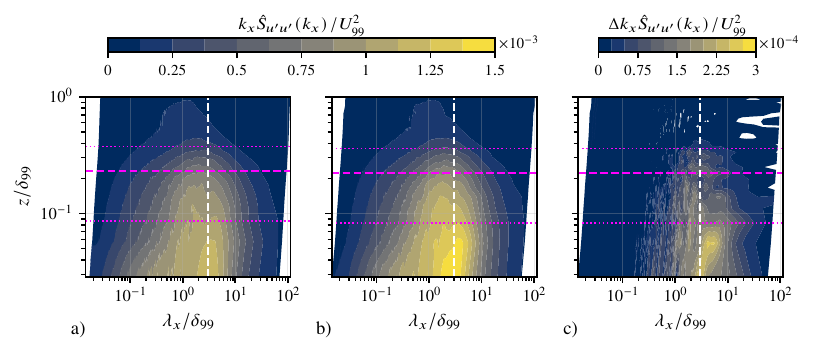}
    \caption{Normalised Spectral energy map for ABL 2 at \(x/D = -1.3\) for a) farm-absent and b) 5x5 farm with \(C_T=0.9\) discs and c) difference. Dashed magenta lines indicate the hub-height and vertical extent of the disc.}
    \label{fig_em_185}
\end{figure}

% ------------------------------------------------------------------ %
Temporal spectra of the boundary layer scans are converted to spatial spectra using Taylor's frozen-turbulence hypothesis with the convection velocity $U_c = 0.82\,U(x,z)$ \citep{Dennis2008, Harun2013}, giving the streamwise wavenumber $k_x = 2\pi f / U_c$ and the corresponding wavelength $\lambda_x = 2\pi/k_x = U_c/f$. The factor $0.82$ accounts for the systematic difference between the local mean velocity and the convection velocity of the energy-containing eddies \citep{Dennis2008}. The correction is applied locally at each $(x,z)$ position using the measured $U(x,z)$, ensuring that comparisons between farm-present and farm-absent spectra are made on a consistent spatial scale basis; using a single hub-height or freestream reference velocity would introduce a systematic shift in $k_x$ between the two configurations as $U$ varies across the induction region. Spectral maps are presented as the pre-multiplied wavenumber spectrum $k_x\hat{S}_{uu}(k_x)/U_{99}^2$, plotted against $\lambda_x/\delta_{99} = 2\pi/(k_x\delta_{99})$ on a logarithmic axis. This representation has the property that equal areas under the curve correspond to equal contributions to $\overline{u'^2}$, making it well suited for identifying the dominant energetic scales. Figures~\ref{fig_em_100} and \ref{fig_em_185} present the pre-multiplied wavenumber spectra as a function of $\lambda_x/\delta_{99}$ and $z/\delta_{99}$, for the farm-present and farm-absent configurations, as well as their difference, of both ABL cases. 

%With $k_x\hat{S}_{uu}(k_x)/U_{99}^2$, which scales by the outer velocity scale.

The most striking feature of the spectral maps is not the farm-induced modification but the structure of the boundary layer turbulence itself. Figures~\ref{fig_em_100}a) and \ref{fig_em_185}a) show the spectragrams (as coined by \cite{Hutchins2007}) of the farm-absent boundary layers. ABL~1 shows the outer peak, centred at \(\lambda_x/\delta_{99} \approx 3\) and \(z/\delta_{99} \approx 0.1\), characteristic of LSM in an APG turbulent boundary layer at moderate Reynolds number \citep{Harun2013}. Indicating the incoming turbulent boundary layer is already experiencing an APG before the influence of the farm. For ABL~2, an outer peak is visible but sits closer to the wall, indicative of a more ZPG boundary layer. Outer peaks location and intensities depend on both \(\mathrm{Re}_{\tau}\) and pressure gradient. For the range of \(\mathrm{Re}_{\tau}\approx 4000-8000\) considered here, we would expect the location of the outer peak to be \(z/\delta_{99} \approx 0.1-0.2\) for APG or \(z/\delta_{99} \approx 0.04-0.06\) for ZPG \citep{Marusic2010, Harun2013}. This likely indicates that ABL~2 is only weakly APG (with \(\partial U/\partial x|_2 < \partial U/\partial x|_1\)). The farm pressure gradients therefore interact with the boundary layers differently. ABL~1 is already an APG boundary layer and becomes more adverse, whereas ABL~2 is only weakly APG becoming adverse. Thus, the resulting boundary layers experienced by the farm are at different stages of development under an APG. The weaker TI amplification of figure~\ref{fig_compare_all_TI} in ABL~2 relative to ABL~1 may therefore partly reflect this history effect rather than the boundary layer depth alone. Despite these differences the two boundary layers are quite similar when expressed in outer scaling ($z/\delta_{99}$), the spectral structure collapses reasonably well across both cases, consistent with the outer-layer similarity arguments of \citet{Townsend1956} and \citet{Bradshaw1967} and reinforcing the conclusion that $\delta_{99}$ is an appropriate outer length scale for characterising the turbulent response.

The large-scale content at $\lambda_x/\delta_{99} \approx 2-3$ is seen to dominate the turbulent kinetic energy across the plotted wall-normal extent of the boundary layer, confirming that the outer-layer large-scale motions carry a disproportionate fraction of the total variance at all heights -- a result consistent with the known energetic dominance of LSMs and VLSMs in ZPG and APG boundary layers \citep{Harun2013}.

The farm-induced modification, when compared between the farm-present and farm-absent maps, is subtle. The overall spectral structure is qualitatively very similar between the two configurations in both ABL cases: the farm does not dramatically restructure the boundary layer turbulence but produces a modest enhancement of the large-scale energy content at $\lambda_x/\delta_{99} \approx 2-3$, consistent with the farm-induced APG acting as a relatively weak additional forcing on a boundary layer whose spectral character is already dominated by outer-layer large-scale motions. This subtlety is physically reasonable: the farm-induced APG in the induction region is mild compared with the strongly non-equilibrium APG configurations studied in the laboratory \citep{Harun2013, Kitsios2016}, and its primary effect is to modestly amplify the existing outer-layer energy peak rather than to create a qualitatively new spectral feature, at wall-normal positions sampled by the disc. Considering figures~\ref{fig_em_100}c) and \ref{fig_em_185}c) the difference between the farm-present and farm-absent cases, we see that both ABLs experience energisation of their outer peaks, but figure~\ref{fig_em_100}c) also shows an increase in energy centred at the hub-height, and across the lower extent of the disc explaining the greater TI-amplification seen in figure~\ref{fig_compare_all_TI}.

\subsection{Hub-height spectra and comparison with canonical inflow models}
%\label{sec:hub_spectra}
% ------------------------------------------------------------------ %

Using the hub-height sweeps, to obtain a farm-scale view of the flow, the spanwise-averaged wavenumber spectrum is computed:
\begin{equation}
    \langle \hat{S}_{uu}(k_x) \rangle_y
    = \frac{1}{y_1 - y_0}
    \int_{y_0}^{y_1} \hat{S}_{uu}(k_x,x,y)\,\mathrm{d}y,
\end{equation}
with units $\text{m}^3\,\text{s}^{-2}\,\text{rad}^{-1}$. The spanwise-averaged pre-multiplied wavenumber spectra at hub height $k_x\langle\hat{S}_{uu}(k_x)\rangle_y/U_{99}^2$ for both ABLs' farm-absent configurations, compare favourably with the \cite{vonKarman1948} and \cite{Kaimal1972} spectral models. These two models represent the most widely used spectral descriptions of atmospheric surface-layer turbulence in wind energy practice \citep{IEC61400}: the von~K\'{a}rm\'{a}n model has a $-5/3$ inertial subrange and a peak set by the integral length scale, whilst the Kaimal model - specified in IEC~61400-1 \citep{IEC61400} - has a broader, lower peak that better represents the spectral shape of neutrally stratified surface-layer turbulence.

ABL~1 and ABL~2 sit between the von~K\'{a}rm\'{a}n and Kaimal models, indicating the incoming boundary layers are an acceptable compromise between canonical BLs and wind industry standards. Both the measured spectra and the canonical von~K\'{a}rm\'{a}n and Kaimal models peak consistently at $\lambda_x/\delta_{99} \approx 3$, the characteristic length scale of the large-scale motions that dominate the outer region of turbulent boundary layers \citep{Hutchins2007, Kim1999}, and both the canonical models and the present laboratory measurements recover this scale when normalised by $\delta_{99}$, confirming that the outer scaling is physically appropriate and that the laboratory boundary layers share the same dominant energetic structure as full-scale atmospheric surface-layer turbulence. This agreement also establishes $\lambda_x/\delta_{99} \approx 2-3$ as the natural reference scale against which the farm-induced spectral modification should be assessed; energy added at this scale can drive coherent, rotor-scale load fluctuations.% most effectively., since $\lambda_x \approx 3\delta_{99}$ is larger than the disc diameter across the range of $\delta_{99}/D$ values studied.

The difference between the farm-present and farm-absent spectra is subtle, as expected given that the spanwise-averaged TI ratio reaches a maximum of only $\approx 1.07$ even for ABL~1. The farm-present spectra carry more energy at scales ($\lambda_x/\delta_{99} \approx 2-3$) characteristic of the LSMs relative to the farm-absent reference. This enhancement is more clear with \(C_T = 0.9\) but less so for \(C_T = 0.7\) and would be difficult to identify from the spectra alone without the supporting evidence of the spanwise-averaged TI results. The hub-height spectra therefore serve primarily to establish the spectral character of the approach flow and its agreement with canonical inflow models. The farm-present and farm-absent spectra are visually very similar, and the hub-height spectra alone are insufficient to fully identify where the modest farm-induced energy increase is concentrated across scales. This motivates the cumulative variance analysis of Section~\ref{sec:spectral_decomp}, which is designed precisely to isolate the scale-selective nature of the farm-induced modification from the background spectral shape.

% ------------------------------------------------------------------ %
\subsection{Spectral variance decomposition: large-scale versus
            small-scale turbulence intensity}
\label{sec:spectral_decomp}
% ------------------------------------------------------------------ %

The cumulative variance in wavenumber space quantifies the fraction of $\overline{u'^2}$ carried by structures with wavenumbers smaller than
$k_x'$ (i.e.\ spatial scales \emph{larger} than $\lambda_x' = 2\pi/k_x'$):
\begin{equation}
\overline{u'^2}(k_x')
= \int_0^{k_x'} \hat{S}_{uu}(k_x)\,\mathrm{d}k_x.
\end{equation}
Normalising by the total variance gives the cumulative distribution function (CDF) of turbulent kinetic energy in wavenumber space:
\begin{equation}
F(k_x')
= \frac{\overline{u^{\prime 2}}(k_x')}{\overline{u^{\prime 2}}_\text{total}},
\qquad F(0) = 0,\quad F(\infty) = 1.
\end{equation}
The normalisation by the total variance of each signal allows direct comparison between signals of different total energy --- since the total variance differs between farm-present and farm-absent configurations by an amount not solely attributable to the farm (e.g. subtle differences in \(U_{\infty}\)), the normalised CDF isolates differences in the \emph{distribution} of energy across scales from differences in total energy content.

The CDF difference is then computed between the farm-present and farm-absent configurations: 
\begin{equation}
    \mathcal{D}(k_x') = F_\text{farm-present}(k_x') - F_\text{farm-absent}(k_x').
\end{equation}
The physical interpretation of $\mathcal{D}(k_x')$ is as follows. At a given wavenumber $k_x'$ (scale $\lambda_x' = 2\pi/k_x'$):
\begin{itemize}
    \item $\mathcal{D}(k_x') > 0$: the presence of the farm has led to an accumulation of a greater fraction of the total energy at scales larger than $\lambda_x'$
    \item $\mathcal{D}(k_x') < 0$: the presence of the farm has led to an accumulation of a greater fraction of the total energy at scales smaller than $\lambda_x'$
    \item $\mathcal{D}(k_x') = 0$: the fractional energy distribution at this scale is unchanged by the presence of the farm.
\end{itemize}
%
%Where $\mathcal{D}$ attains its maximum positive value at $k_x^*$, the corresponding scale $\lambda_x^* = 2\pi/k_x^*$ is the scale at which the farm most strongly redistributes energy toward larger structures --- the scale that ``gains most'' from the farm-induced flow modification. 

\begin{figure}[ht] % Figure 8
    \centering
    \includegraphics[width=13cm]{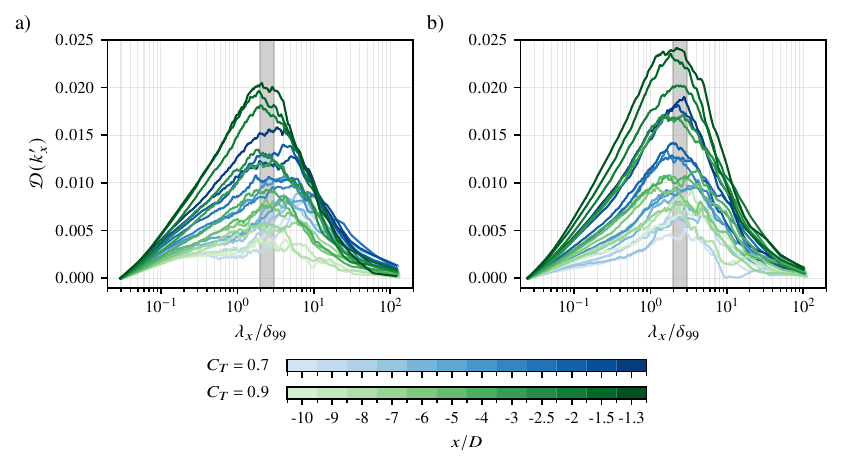}
    \caption{The CDF difference plotted against normalised wavelength for a) ABL~1 and  b) ABL~2 at the hub height at \(x/D = -1.3\) for a \(5\times 5\) farm. The shaded region is \(\lambda_x/\delta_{99}  = 2-3\).}
    \label{fig_cum_sum_100}
\end{figure}

Figure~\ref{fig_cum_sum_100} presents the normalised CDF difference as a function of $\lambda_x/\delta_{99}$ for the farm-present and farm-absent configurations of ABL~1 and ABL~2 at hub height. The CDF difference shows a clear, localised spike in the rate of variance accumulation in the range $\lambda_x/\delta_{99} \approx 2$--$3$, confirming that the farm-induced variance increase is preferentially occurring at LSM and VLSM wavenumbers \citep{Hutchins2007, Kim1999}, consistent with the outer-layer amplification driven by the farm-induced adverse pressure gradient. At $\lambda_x/\delta_{99} \ll 1$ the difference in cumulative variance is small, indicating that the small-scale turbulence energy is only weakly modified by the farm. As \(\mathcal{D} > 0 \) for all \(k_x\) shows that the turbulence amplification is purely additive and no scales are suppressed in the process. 

\begin{figure}[ht] % Figure 9
    \centering
    \includegraphics[width=10cm]{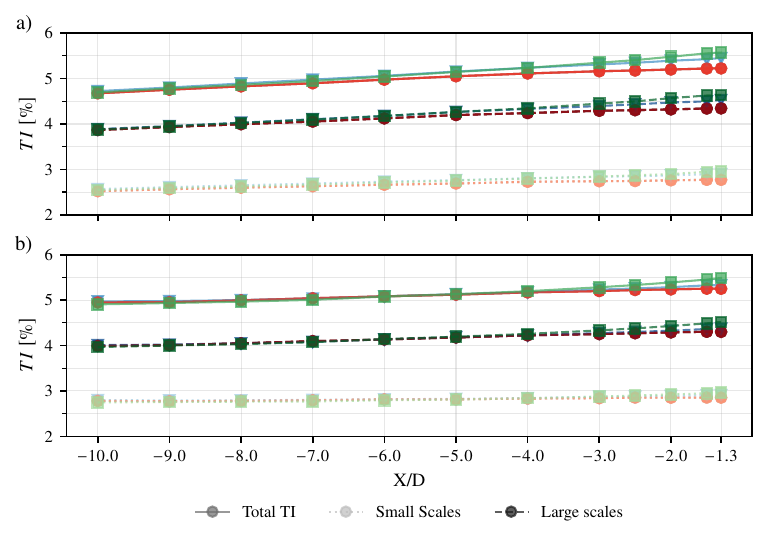}
    \caption{The turbulence decomposition into the large scales, darker lines, and the smaller scales, the lighter lines. For a) ABL 1 and b) ABL 2. Larger scales are scales larger than the boundary layer depth, \(\lambda_x/\delta_{99}\). \(C_T = 0.7\) in shades of blue, \(C_T = 0.9\) in shades of green, and farm-absent in shades of red.}
    \label{fig_TI_freq_1}
\end{figure}

To quantify the scale dependence of the farm-induced turbulence amplification, the streamwise turbulent kinetic energy is decomposed into large-scale and small-scale contributions separated at the threshold $\lambda_x / \delta_{99} = 2\pi/(k_c \delta_{99}) = 1$, i.e. scales larger than the boundary layer are large. Scales with $\lambda_x > \delta_{99}$ are associated with LSMs and VLSMs whose wall-normal extent is comparable to the boundary layer depth and whose amplification is the expected signature of an APG outer-layer response \citep{Harun2013, Hutchins2007}. Scales with $\lambda_x < \delta_{99}$ are associated with the near-wall cycle and the inertial subrange, which are expected to respond less directly to the farm-scale pressure gradient. The large-scale and small-scale turbulence intensities are defined respectively as:
\begin{equation}
    \begin{aligned}
        \mathrm{TI}_\text{LS}
        &= \frac{1}{U_\text{ref}}
           \left(\int_{0}^{k_c}
                 \hat{S}_{uu}(k_x)\,\mathrm{d}k_x\right)^{1/2},
        &
        \mathrm{TI}_\text{SS}
        &= \frac{1}{U_\text{ref}}
           \left(\int_{k_c}^{\infty}
                 \hat{S}_{uu}(k_x)\,\mathrm{d}k_x\right)^{1/2}.
    \end{aligned}
    %\label{eq:TI_LS_SS}
\end{equation}
where the integrals are evaluated numerically from the measured spectra and $\mathrm{TI}_\mathrm{LS}^2 + \mathrm{TI}_\mathrm{SS}^2 = \mathrm{TI}^2$. The large-scale and small-scale \(\mathrm{TI}\) decomposition is shown in Figure~\ref{fig_TI_freq_1} as a function of streamwise position for both ABL cases. The large-scales $\mathrm{TI}_{\text{LS},\text{farm}}$ mirrors the spatial development of the total $\mathrm{TI}$ ratio in Figure~\ref{fig_compare_all_TI} and accounts for the majority of the total \(\mathrm{TI}\) amplification in both cases. The small-scales $TI_{\text{SS},\text{farm}}$ remain almost unchanged throughout the induction region, confirming that the small-scale turbulence responds weakly to the farm-induced pressure gradient over the streamwise distances considered. The large-to-small scale TI ratio increases progressively as the farm is approached, indicating that the preferential large-scale energisation becomes more pronounced as the adverse pressure gradient strengthens and the boundary layer moves further from ZPG equilibrium; consistent with the DNS results of \citet{Kitsios2016}, who showed that the outer-layer peak in $\overline{u'^2}$ grows relative to the near-wall peak (which remains unchanged) as APG increases. 

The physical interpretation is clear: the farm-induced adverse pressure gradient drives the turbulent boundary layer into an increasingly non-equilibrium APG state, in which energy is preferentially transferred from the mean flow to the large-scale outer-layer motions through the modification of the wall-normal shear production \citep{Harun2013}. The small scales, sustained primarily by the near-wall production cycle and the inertial energy cascade, are only weakly affected. The result is a selective amplification of the turbulence at wavenumbers corresponding to scales comparable to the boundary layer depth and by extension the rotor diameter --- precisely the scales most consequential for wind turbine performance and fatigue loading. The \(\mathrm{TI}\) increase documented in Section~\ref{sec:spanwise} is therefore not a broadband effect, but a large-scale phenomenon with a specific wavenumber signature that current inflow turbulence models, which prescribe the full spectrum through canonical von~K\'{a}rm\'{a}n or Kaimal models, may underestimate.

%The spectral decomposition threshold \(f_c = 0.82 \overline{U}_{\mathrm{farm-absent}}/ \delta_{99, \ \mathrm{farm-absent}}\) is evaluated using the farm-absent reference boundary layer depth at each streamwise station, ensuring that the large-scale and small-scale bands are defined with respect to the undisturbed outer length scale. This choice is preferred over using the farm-present \(\delta_{99}\) because the latter is itself modified by the farm-induced pressure gradient, which would conflate the thickening of the boundary layer with the spectral redistribution of energy — the two effects the decomposition is designed to separate.

% ================================================================== %
\section{Conclusions}
% ================================================================== %

This work investigated the upstream induction region of a model wind farm and its interaction with the turbulent boundary layer, using an array of porous discs and two boundary layers of different depths, with farm-present and farm-absent configurations compared throughout. Mean velocity blockage and turbulence intensity amplification rank consistently across the two boundary layers, both larger for the shallower ABL~1. For the mean flow this is consistent with reduced vertical bypass in a shallower layer \citep{Ivanell2026, Nishino2020}; for the turbulence, however, ABL~1 and ABL~2 differ in pressure-gradient history as well as depth, and the turbulent memory effect \citep{Bobke2017, SanmiguelVila2017} means we cannot separate the contribution of $\delta_{99}/D$ from that of history. Farm geometry has little influence: depth beyond three rows and row spacing barely affect either response, whereas the disc thrust coefficient $C_T$ has a strong effect on both, identifying it as the dominant control parameter and suggesting that derating the leading rows may be operationally relevant. The spectral response separates into a fixed location and a variable magnitude. The farm-induced turbulence increase is carried predominately by large scales, peaking at $\lambda_x/\delta_{99} \approx 2-3$ --- consistent with APG theory \citep{Harun2013, Kitsios2016} and with canonical atmospheric spectral models --- and this peak location is essentially invariant across both boundary layer case and farm configuration. We interpret this as the farm acting as a broadband pressure forcing while the boundary layer responds at its own natural receptive scale, consistent with outer-layer similarity \citep{Townsend1956, Bradshaw1967, Hutchins2012}. The magnitude of the response, unlike its location, scales with increasing $C_T$: the farm sets the forcing amplitude, the boundary layer sets the scale at which it is expressed. Because the amplified scales exceed $\delta_{99}$, this is of practical concern, as such scales drive coherent, whole-rotor loading most damaging under fatigue \citep{Thomsen1999, Sutherland1999, Nejad2014}. Applying a lab-scale canonical boundary layer framework to a true atmospheric boundary layer is supported by \citet{Hutchins2012}, who established equivalence between atmospheric and laboratory large-scale structure under neutral stratification. These results establish the induction region as a genuine, physically coherent APG boundary layer problem, with scale-selective turbulence modification not captured by canonical inflow turbulence models. The porous disc abstraction is necessarily open-loop, suppressing the feedback by which amplified turbulence would modify effective farm thrust and hence the upstream pressure gradient; closing this loop with dynamically responsive rotor models remains for future work. Boundary layer depth, $\delta_{99}/D$, emerges as a candidate key parameter for both the mean and turbulent response, though its role in the turbulence response cannot yet be separated from pressure-gradient history --- disentangling the two is a natural target for future work.

\begin{Backmatter}

\paragraph{Acknowledgements}
We gratefully acknowledge the advice of F. J. G. de Oliveira, Y. Pan, and M. Bourhis during the conceptualisation of this work and R. Harbison, I. W. James, and  A. D. Smith for assistance during the experimental campaigns.

\paragraph{Funding Statement}
A.T. McGlade and O.R.H. Buxton gratefully acknowledge the financial support provided by the Engineering and Physical Sciences Research Council (EPSRC), through grant no. EP/V006436/1.

\paragraph{Declaration of Interests}
The authors declare no conflict of interest.

\paragraph{Data Availability Statement}
Raw data are available from the corresponding author (A.T. McGlade).

\paragraph{Ethical Standards}
The research meets all ethical guidelines, including adherence to the legal requirements of the study country.

\end{Backmatter}

\end{document}